\documentclass[9pt,conference]{IEEEtran}
\usepackage[preprint]{dcase2025}
\usepackage{bm}

\usepackage{amsmath,graphicx,url,times,booktabs,tabularx}
\usepackage{siunitx}

\title{STEREO SOUND EVENT LOCALIZATION AND DETECTION WITH ONSCREEN/OFFSCREEN CLASSIFICATION}

\name{Kazuki Shimada$^{1}$,
      Archontis Politis$^{2}$,
      Iran R. Roman$^{3}$,
      Parthasaarathy Sudarsanam$^{2}$,
      David Diaz-Guerra$^{2}$,
      }
\secondlinename{	  
      Ruchi Pandey$^{2}$,
      Kengo Uchida$^{1}$,
      Yuichiro Koyama$^{4}$,
      Naoya Takahashi$^{5}$,
      Takashi Shibuya$^{1}$,
      }
\thirdlinename{
      Shusuke Takahashi$^{4}$,
      Tuomas Virtanen$^{2}$,
      Yuki Mitsufuji$^{6,7}$
      }

\address{$^1$ Sony AI, Japan,
         $^2$ Audio Research Group, Tampere University, Finland,
         $^3$ Queen Mary University of London, UK \\
         $^4$ Sony Group Corporation, Japan,
         $^5$ Sony AI, Switzerland,
         $^6$ Sony AI, USA,
         $^7$ Sony Group Corporation, USA
         }

\begin{document}

\maketitle

\begin{abstract}

This paper presents the objective, dataset, baseline, and metrics of Task 3 of the DCASE2025 Challenge on sound event localization and detection (SELD).
In previous editions, the challenge used four-channel audio formats of first-order Ambisonics (FOA) and microphone array.
In contrast, this year's challenge investigates SELD with stereo audio data (termed stereo SELD).
This change shifts the focus from more specialized 360° audio and audiovisual scene analysis to more commonplace audio and media scenarios with limited field-of-view (FOV).
Due to inherent angular ambiguities in stereo audio data, the task focuses on direction-of-arrival (DOA) estimation in the azimuth plane (left-right axis) along with distance estimation.
The challenge remains divided into two tracks: audio-only and audiovisual, with the audiovisual track introducing a new sub-task of onscreen/offscreen event classification necessitated by the limited FOV.
This challenge introduces the DCASE2025 Task3 Stereo SELD Dataset, whose stereo audio and perspective video clips are sampled and converted from the STARSS23 recordings.
The baseline system is designed to process stereo audio and corresponding video frames as inputs.
In addition to the typical SELD event classification and localization, it integrates onscreen/offscreen classification for the audiovisual track.
The evaluation metrics have been modified to introduce an onscreen/offscreen accuracy metric, which assesses the models' ability to identify which sound sources are onscreen.
In the experimental evaluation, the baseline system performs reasonably well with the stereo audio data.
\end{abstract}


\section{Introduction}
\label{sec:intro}

Sound event localization and detection (SELD) involves identifying sound events, tracking their temporal occurrence, and estimating their spatial direction or position from multichannel audio inputs.
This results in a spatiotemporal characterization of the acoustic scene that can be used in a wide range of machine cognition tasks, such as inference on the type of environment, self-localization, navigation with visually occluded targets, tracking of specific types of sound sources, smart-home applications, scene visualization systems, and acoustic monitoring, among others.
SELD has received increasing attention since the earliest publications~\cite{adavanne2018sound}.
One of the most significant research efforts on this topic has been centered around the DCASE challenge\footnote{\url{https://dcase.community/challenge2025/}}, with the task developing every year in terms of data complexity, data realism, or task complexity.

The first three iterations of the task (2019--2021)~\cite{adavanne2019multi,politis2020dataset,politis2021dataset,politis2020overview} were based on synthetic multichannel audio data that included multichannel ambient noise and reverberation.
The data were provided in two types of four-channel audio formats, i.e., first-order Ambisonics (FOA) and tetrahedral microphone array.
They were synthesized using multi-room and multi-point spatial room impulse responses (SRIRs) from real spaces, which enabled the synthesis of both static and moving reverberant sound events.
The first three challenges considered moving sound events, non-target interfering directional sound events, and overlapping events of the same class.
The top-ranked systems in the three challenges excelled at addressing these issues by employing effective audio features~\cite{nguyen2022salsa}, advanced data augmentation strategies~\cite{mazzon2019first,wang2023four,koyama2022spatial}, or improved SELD frameworks~\cite{kapka2019sound,cao2019polyphonic,nguyen2020sequence,cao2021improved,shimada2021accdoa}.

Although those synthetic datasets carefully emulated the acoustical and spatial properties of sound scenes, they lacked some important aspects of real sound scenes, mainly the natural temporal and spatial occurrences and co-occurrences that characterize real sound events, resulting from the scene environment and the actions and interactions of the agents within it.
To advance SELD research in that direction, the following three iterations (2022--2024)~\cite{politisstarss22,shimada2023starss23,aparicio2024baseline} were based on a new dataset of spatial recordings of real scenes.
The original 32-channel recordings were converted to the two types of four-channel audio formats (i.e., FOA and microphone array) and made available to participants.
Annotations of sound event activities for 13 classes were compiled by human listeners and combined with optical tracking data of the source positions that generated those sound events.
Eleven hours of such material were collected in multiple rooms in two different sites.
In contrast to the class-balanced synthetic datasets of the earlier studies~\cite{adavanne2019multi,politis2020dataset,politis2021dataset,politis2020overview}, the real recordings featured highly unbalanced class distributions.
To cope with the increased difficulty of the task and the limited amount of training data, participants were allowed to use external data, additional simulations of recordings, and pre-trained models.
Additionally, from the DCASE2023 Challenge~\cite{shimada2023starss23}, the 360° video recordings were provided to stimulate further cross-modal developments in SELD research.
The full field-of-view (FOV) video data are temporally synchronized and spatially aligned with the audio recordings.
The effective integration of audiovisual data~\cite{Du_NERCSLIP_dcase23task3_report,berghi2024fusion}, the efficient use of external resources~\cite{Du_NERCSLIP_dcase22task3_report,Hu_IACAS_dcase22task3_report}, and more powerful architectures driven by attention mechanisms~\cite{partha2021dcase,Du_NERCSLIP_dcase24task3_report} enabled the top participants to achieve competitive results with substantial gains over the baselines.

While the previous iterations used full FOV four-channel audio formats, this year's task\footnote{\url{https://dcase.community/challenge2025/task-stereo-sound-event-localization-and-detection-in-regular-video-content}} investigates SELD with stereo audio data (called stereo SELD).
This change aims to reflect commonplace audio and media scenarios.
Spatial analysis of stereo audio data can be applied to a wide variety of media content~\cite{shimada2024savgbench} since it is the default audio format for media capture or production.
As the stereo audio format used in the task exhibits angular ambiguity in the top-bottom and front-back directions, the task focuses on direction-of-arrival (DOA) estimation of azimuth angles only along the left-right axis, together with source distance estimation, which has been explored in prior SELD and spatial audio studies~\cite{krause2024sound,krause2023binaural,kushwaha2023sound,liang2023reconstructing}.
This challenge remains divided into two tracks: evaluation of stereo SELD models with audio-only input (Track A) or audiovisual input (Track B).
Perspective video data are employed to reflect widely used media formats.
Due to the limited FOV, the audiovisual track poses a new sub-task: onscreen/offscreen classification of the detected events.
Such prediction can be helpful in further audiovisual processing, such as audio enhancement of objects contained in the video FOV or automated steering of cameras towards target sources outside the FOV.

Adapting the challenge for the stereo SELD task requires modifications at multiple fronts.
\begin{itemize}
    \item A new DCASE2025 Task3 Stereo SELD Dataset\footnote{\url{https://zenodo.org/records/15559774}} derived from the STARSS23 dataset~\cite{politisstarss22,shimada2023starss23} extracting limited FOV perspective videos from the omnidirectional 360° videos, and corresponding stereo tracks oriented at the center of those videos from the FOA audio format.
    \item A baseline system\footnote{\url{https://github.com/partha2409/DCASE2025_seld_baseline}} in an audio-only and audiovisual configuration that processes stereo audio and corresponding video frames as inputs. It outputs sound event labels and horizontal DOAs and distances over time without front or back distinction. The audiovisual baseline additionally outputs an onscreen/offscreen binary label for each detected event.
    \item The evaluation metrics are adapted to these task changes, with horizontal-only DOA localization errors and an onscreen/offscreen accuracy metric, which assesses the models' ability to identify which sound sources are onscreen.
\end{itemize}

\section{Stereo SELD Dataset}
\label{sec:data}

\subsection{Overview}
\label{ssec:data-overview}

This challenge uses the DCASE2025 Task3 Stereo SELD Dataset, which comprises stereo audio and perspective video data, derived from the STARSS23 dataset~\cite{politisstarss22,shimada2023starss23}.
STARSS23 contains multichannel audio (e.g., FOA) and 360° video recordings of sound scenes in various rooms and environments, together with temporal and spatial annotations of target event classes.
The STARSS23's FOA audio and 360° video data have been converted to stereo audio and perspective video data, simulating regular media content.
The data construction code for the task is made available\footnote{\url{https://github.com/SonyResearch/dcase2025_stereo_seld_data_generator}}.

\begin{figure}[t]
    \centering
    \centerline{\includegraphics[width=0.99\linewidth]{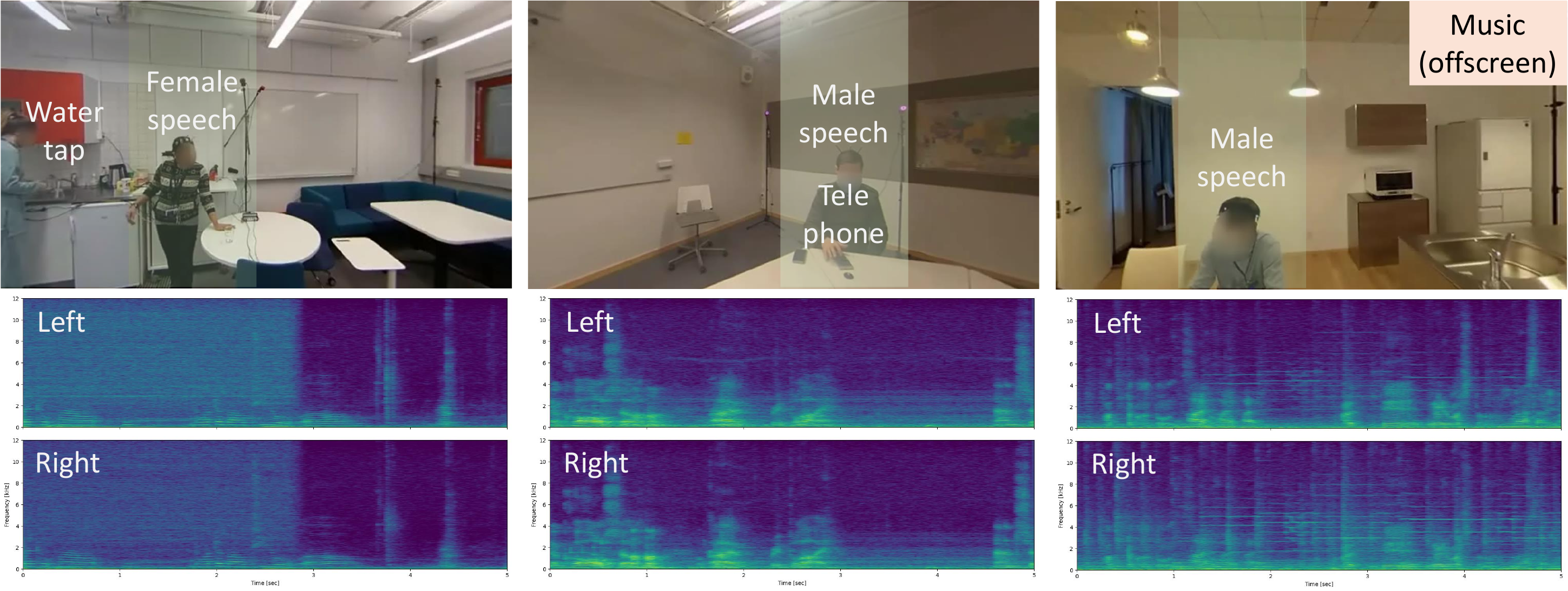}}
    \caption{Examples of the DCASE2025 Task3 Stereo SELD Dataset. In the clip on the left, water is coming out of a faucet on the left, and a woman is speaking from the left of the center. In the middle, from the right of the center, a phone rings, and a man speaks. In the clip on the right, a man is speaking from the left of the center, and music is playing offscreen (from the right).}
    \label{fig:dataset}
\end{figure}

The DCASE2025 Task3 Stereo SELD Dataset is split into a development dataset and an evaluation dataset.
The development dataset comprises 30,000 clips of 5-second duration, with a total time of 41.7 hours.
The evaluation dataset contains 10,000 clips of 5-second duration, totaling 13.9 hours.
We release the development dataset to the public and keep the evaluation one for the challenge evaluation.
We provide a training-testing split for consistent reporting of results across systems on the development dataset.
The DCASE2025 Task3 Stereo SELD Dataset inherits the 13 event classes from STARSS23, including speech, telephone, and water tap.
The audio data are formatted as stereo and have a sampling rate of 24 kHz.
The video resolution (width:height) is set to 640:360 pixels, with an aspect ratio of 16:9, widely used in media content.
The horizontal FOV is set to 100°.
The video frames per second (fps) rate is 29.97.
Examples of the dataset are shown in Fig~\ref{fig:dataset}.

\subsection{Data Construction}
\label{ssec:data-const}

The DCASE2025 Task3 Stereo SELD Dataset is constructed as follows.
We first sample 5-second clips from the STARSS23 recordings.
Then, we convert the 5-second FOA audio and 360° video to generate stereo audio and perspective video data corresponding to a fixed point of view.
According to the fixed viewing angle, we first rotate the FOA audio~\cite{mazzon2019first}.
Then, we convert the rotated FOA audio to stereo audio, emulating a mid-side (M/S) recording technique~\cite{wilkins2023two}.
We convert the 360° video to a perspective video with the same viewing angle as the audio, using a python library\footnote{\url{https://github.com/sunset1995/py360convert}}.
We also rotate the STARSS23's DOA labels to new DOA labels centered at the fixed viewing angle.
Rotated azimuth labels that point to the back hemisphere are folded from back to front, considering the front-back ambiguity.
The elevation labels are omitted due to top-bottom ambiguity and weak capability of stereo formats to resolve elevation.
The distance labels are kept the same as the STARSS23 ones.
To get the binary onscreen/offscreen event labels, we compare the new DOA labels with the azimuthal range of the FOV in the perspective video.

The details of the sampling procedure are provided below.
For each sampling, we first randomly select a STARSS23 recording file.
We employ duration-weighted random choice, i.e., each file has a probability of being selected proportional to its duration.
This aims to take a large number of samples from files of long duration.
Then, we select a start frame uniformly within each file.
We select a horizontal viewing angle uniformly over 360° while keeping the vertical viewing angle at 0° elevation.
We repeat the sampling procedure until the desired number of clips have been sampled.
We do not control explicitly for the presence of target sound events in the randomly sampled clips.
Hence, there is a small amount of clips that do not contain any target sound event.
We do not consider that to be an issue since these clips provide confirmation clips for false positives.
The class distribution across all frames after random sampling is similar to that of STARSS23~\cite{politisstarss22,shimada2023starss23}.
The onscreen/offscreen distribution across all frames is approximately 1:3.

\subsection{Audio Format Description}
\label{ssec:data-format}

The stereo audio format is derived from the full-sphere FOA format used in the previous iterations of the challenge.
For recording details and encoding specifications of the FOA format, refer to the previous task description papers~\cite{adavanne2019multi,politis2020dataset,politis2021dataset,politisstarss22}.
The stereo format specification emulates a coincident M/S stereo recording technique using cardioid microphones pointing fully at 90° (left channel) and -90° (right channel).
Since any M/S stereo recording configuration can be extracted from FOA signals, in this simple M/S case, they are derived using only the omnidirectional and left-right dipole components of the FOA signals.
More specifically, for FOA signals following an ACN/SN3D convention ordered as $[W(n),Y(n),Z(n),X(n)]$, the stereo signals $[L(n),R(n)]$ are then simply:
\begin{align}
    L(n) = W(n) + Y(n), \\
    R(n) = W(n) - Y(n),
\end{align}
where $W(n)$ is the zeroth-order omnidirectional signal of the FOA audio, and $Y(n)$ is the first-order horizontal (left-right) dipole component~\cite{wilkins2023two}.

\section{Baseline Model}
\label{sec:baseline}

\begin{figure}[t]
   \centering
	{{\includegraphics[trim={1.7cm 0.2cm 4.8cm 0.3cm},clip,width=0.99\linewidth]{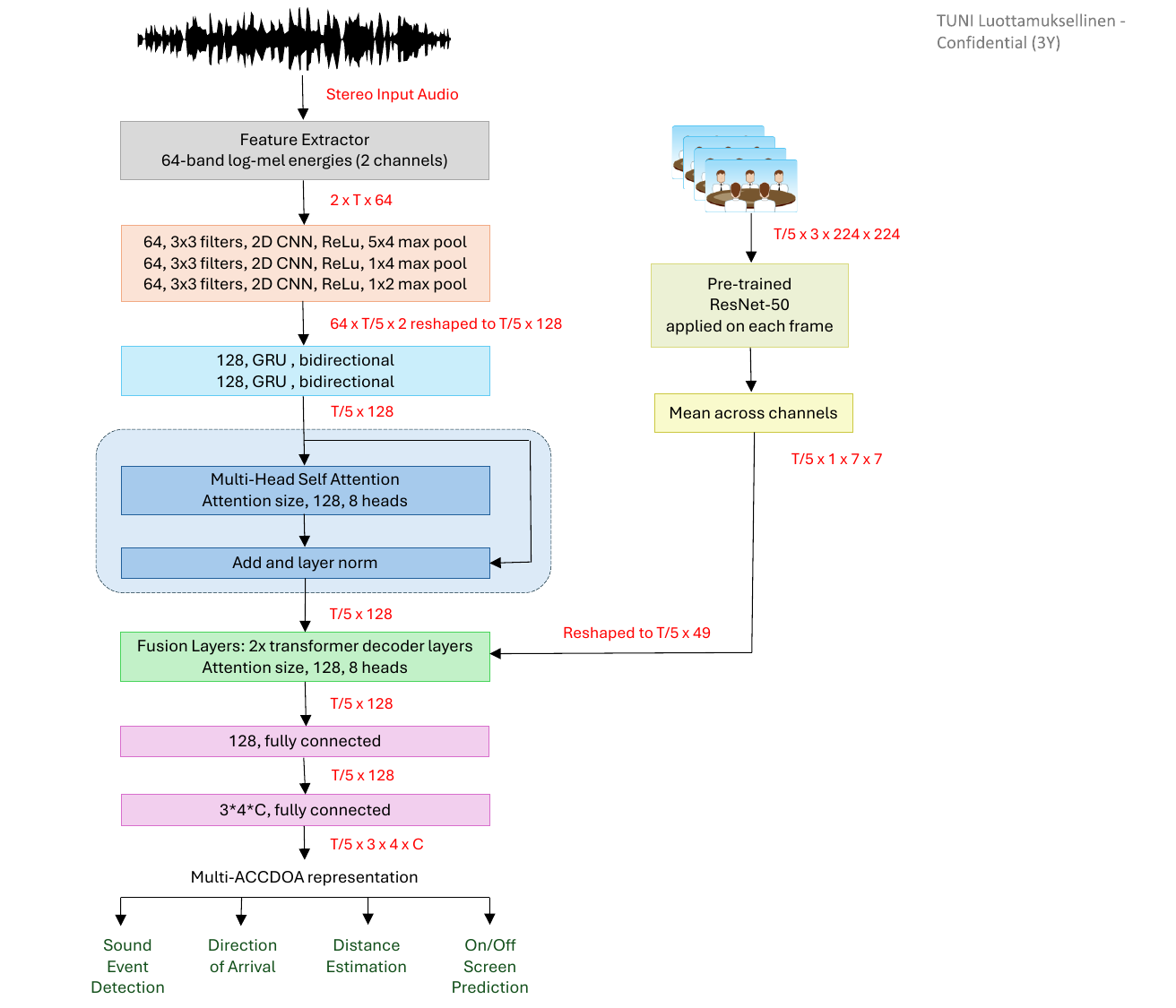}}}
	\caption{Audiovisual baseline model architecture for the stereo SELD task.}
	\label{fig:baseline}
\end{figure}

Figure~\ref{fig:baseline} shows the baseline architecture for the stereo SELD task that processes stereo audio and corresponding video frames as inputs. For the audio-only track, we consider the same architecture from the previous edition~\cite{aparicio2024baseline} and for the audiovisual track, onscreen/offscreen classification is integrated into the baseline architecture. For the audio-only track, 64-band log-mel spectrograms are extracted from stereo audio and fed to a convolutional recurrent neural network (CRNN). It consists of three convolutional layers (with 64 filters each), bidirectional gated recurrent unit (GRU) layers, and multi-head self-attention blocks with an attention size of 128 and eight attention heads~\cite{aparicio2024baseline,partha2021dcase}. For the audiovisual baseline, a pre-trained ResNet-50 model~\cite{he2016deep} is used to extract visual features at a frame rate of 10 fps. These features are reshaped and combined with audio features through transformer decoder-based fusion layers utilizing cross-attention mechanisms~\cite{vaswani2017attention,berghi2024fusion}. The fused audiovisual representations are then passed through fully connected layers to obtain final predictions.

The model follows the multi-ACCDOA output format~\cite{shimada2022multi,krause2024sound}, predicting up to three simultaneous sound events per class at each time frame. Each sound event class in the multi-ACCDOA output is represented by three regressors that estimate the Cartesian $[x, y]$ coordinates of the azimuth DOA around the microphone and a distance value. In case of the audiovisual model, there is an additional binary output neuron that predicts whether the sound event is within the video frame or outside of it. The model is trained to minimize mean squared error loss for the DOA and distance predictions, and binary cross-entropy loss for the onscreen/offscreen classification. This baseline architecture offers participants a comprehensive foundation for exploring and advancing SELD techniques suitable for realistic stereo audiovisual contexts.

\section{Synthetic Data to Support Model Training}
\label{sec:synthetic_data}

To enhance the robustness and generalization of the baseline models, we incorporate synthetic data alongside the development dataset clips. These synthetic examples are generated using publicly available tools designed to simulate realistic spatial sound scenes and video content aligned with the task’s data format.

We use the SpatialScaper library~\cite{roman2024spatial}, which generates FOA audio by convolving isolated sound samples of known sound types with SRIRs. It additionally generates metadata of the respective annotated sound events including their sound class, onset, offsets, and location. We utilize a curated subset of FSD50K~\cite{fonseca2021fsd50k} samples that match the target classes of the challenge. The music class is sampled from the FMA dataset~\cite{defferrard2017fma}. The SpatialScaper scripts and documentation are available in their official repository\footnote{\url{https://github.com/iranroman/SpatialScaper}}. Synthetic 360° videos are generated using stock background panoramas and object/person overlays. These visual elements are placed and animated based on the spatial coordinates and timing information from the synthetic SpatialScaper metadata. The code to generate this video is hosted by the SELDVisualSynth project~\cite{roman2025generating} on github\footnote{\url{https://github.com/adrianSRoman/SELDVisualSynth}}.

To create training-ready clips from the synthetic data, we apply the same stereo SELD data generation pipeline used for the official real-world dataset. This pipeline converts synthetic FOA audio and 360° video into stereo audio and perspective video based on a specified viewing angle, and prepares metadata accordingly. 


\section{Evaluation Metrics}
\label{sec:eval_metric}

\begin{table*}[t]
    \centering
    \caption{
    SELD performance of the audio-only and audiovisual baseline models on the \textbf{development} dataset of the DCASE2025 Task3 Stereo SELD Dataset.
    }
    \scalebox{1.00}{
        \begin{tabular}{l|ccccc}
        \toprule
         & Macro $\mathrm{F}_{20^{\circ}/1}$ $(\%)$ & Macro $\mathrm{F}_{20^{\circ}/1/\mathrm{onoff}}$ $(\%)$ & $\mathrm{DOAE}_\mathrm{CD}$ $({}^\circ)$ & $\mathrm{RDE}_\mathrm{CD}$ $(\%)$ & Onscreen/offscreen accuracy $(\%)$ \\
        \midrule
        Audio-only & 22.8 & N/A & 24.5 & 41.0 & N/A \\
        Audiovisual & 26.8 & 20.0 & 23.8 & 40.0 & 80.0 \\
        \bottomrule
        \end{tabular}
        }
    \label{tab:baseline-dev}
\end{table*}

\begin{table*}[t]
    \centering
    \caption{
    SELD performance of the audio-only and audiovisual baseline models on the \textbf{evaluation} dataset of the DCASE2025 Task3 Stereo SELD Dataset.
    }
    \scalebox{1.00}{
        \begin{tabular}{l|ccccc}
        \toprule
         & Macro $\mathrm{F}_{20^{\circ}/1}$ $(\%)$ & Macro $\mathrm{F}_{20^{\circ}/1/\mathrm{onoff}}$ $(\%)$ & $\mathrm{DOAE}_\mathrm{CD}$ $({}^\circ)$ & $\mathrm{RDE}_\mathrm{CD}$ $(\%)$ & Onscreen/offscreen accuracy $(\%)$ \\
        \midrule
        Audio-only & 26.1 & N/A & 23.0 & 32.3 & N/A \\
        Audiovisual & 27.5 & 20.8 & 22.2 & 37.7 & 77.8 \\
        \bottomrule
        \end{tabular}
        }
    \label{tab:baseline-eval}
\end{table*}

Following last-year decision of reducing the number and complexity of the evaluation metrics, this year baseline and submitted models have been evaluated according to their localization-dependent F-score ($\mathrm{F}_{20^{\circ}/1}$) and their classification-dependent DOA estimation error ($\mathrm{DOAE}_\mathrm{CD}$) and relative distance estimation error ($\mathrm{RDE}_\mathrm{CD}$) computed at frame level. Apart form last-year metrics, this year we have introduced an onscreen/offscreen accuracy metric to evaluate the ability of the models to identify which sound sources are onscreen.

As in previous editions, we compute the classification-dependent localization metrics taking into account only those detections correctly classified, and for the localization-dependent F-score we only accept as true positives those detections whose DOA and distance estimations are below a spatial threshold. Additionally in this challenge, on the audiovisual track, we also require a correct onscreen/offscreen estimation in order to accept estimates as true positives ($\mathrm{F}_{20^{\circ}/1/\mathrm{onoff}}$). As last year, the DOA estimation error threshold is $\ang{20}$ and the relative distance estimation error threshold is 1.

The main difference in the evaluation compared to previous editions of the challenge is in the way we rank the submissions. Since SELD models have to perform a combination of tasks whose various aspects are usually evaluated using different metrics, the global ranking in the previous years was decided by combining the individual rankings of every metric. This strategy seemed to work reasonably well when the number of tasks was limited (sound event detection, classification, and DOA estimation). However, last year, the distance estimation task was additionally introduced and it proved to be a much more challenging one, with the differences between all submitted models being quite narrow. In that case, we observed that combination of rankings based on all metrics can lead to some undesired artifacts. Models with very minor improvements in distance estimation could scale up many positions in the relative distance estimation error ranking, and that could have a very big impact in the final ranking, overtaking models that clearly obtained better results in the other metrics. 
To avoid this problem, this year, the global ranking is done exclusively according to the localization-dependent F-score. Obtaining a good result in this metric requires a good localization performance due to the spatial threshold and, in for the audiovisual track, we also require a correct onscreen/offscreen estimation, so the localization-dependent F-score effectively evaluates all the SELD tasks.

\section{Baseline Results}
\label{sec:results}

The evaluation metric scores of the baseline models on the development and evaluation datasets are given in Table~\ref{tab:baseline-dev} and \ref{tab:baseline-eval}, respectively. While this challenge uses the localization-aware F1-score considering spatial and onscreen thresholds for ranking in the audiovisual track, the F1-score considering only spatial thresholds is also shown for comparing the audiovisual result to the audio-only result.


We can observe how the audiovisual model slightly outperforms the audio-only model in the detection and angular localization metrics, but obtain a worse result for its distance estimation on the evaluation dataset. This shows how integrating the visual information in the SELD problem is still a challenging task even when the audio is reduced to a stereo signal and the image is simplified into a conventional rectangular frame.

\section{Dataset bias}
\label{sec:data_bias}

\begin{table}[t]
    \centering
    \caption{SELD performance of the models in the evaluation dataset replacing the estimated distance by the average distance of the estimated class on the development dataset.}
    \scalebox{0.84}{
    \begin{tabular}{l|ccc}
    \toprule
     & Macro $\mathrm{F}_{20^{\circ}/1}$ $(\%)$ & Macro $\mathrm{F}_{20^{\circ}/1/\mathrm{onoff}}$ $(\%)$ & $\mathrm{RDE}_\mathrm{CD}$ $(\%)$ \\
     \midrule
    Perfect detector & 98.0 & 98.0 & 30.8 \\
    AO baseline & 26.5 & N/A & 30.3 \\
    AV baseline & 27.9 & 21.1 & 33.5 \\
    \bottomrule
    \end{tabular}
    }
    \label{tab:bias}
\end{table}

The proposed dataset contains several biases that can be exploited to perform some estimations without even extracting any information from the audio or audiovisual signals. The most simple example is the onscreen/offscreen estimation. Since the viewing angle was chosen randomly for every clip regardless of their content, \SI{77.5}{\percent} of the sound sources are offscreen. This shows that the onscreen/offscreen accuracy of the audiovisual baseline is around a random guess adjusted to the data distribution, which again points out that the model is struggling to learn how to exploit the visual information.

The bias on the distance of the sources is a bit more complex since it depends on the class of the audio event. To analyze it, we computed the average distance of the sources of every class in the training set, but the ability of the models to exploit this bias depends on their ability to obtain a correct class estimation. Table~\ref{tab:bias} shows the distance-dependent metrics for the evaluation dataset that result from replacing the distance estimation by the average distance of the estimated class for a perfect detector and for the two baselines.


We can see how the distance estimation of the baseline models is worse than what they could have achieved by just learning the bias of the dataset, which showcases the difficulty of estimating the distance of a sound source using only stereo audio. We can also observe how the distance estimation has a little impact on the localization-dependent F-score due to the large RDE threshold used when computing it.

\section{Conclusion}
\label{sec:conclu}


We introduced the stereo SELD task in the DCASE2025 Challenge, including a new audiovisual sub-task focused on onscreen/offscreen event classification. The proposed dataset and baseline demonstrate that stereo SELD is feasible, with initial results indicating promise for practical applications. However, the audiovisual track remains challenging; our baseline suggests that current models underutilize visual information, leaving significant room for improvement. Similarly, distance estimation is still unreliable and warrants further exploration. We hope the community’s participation will drive progress in tackling these challenges and lead to more robust and ecologically valid SELD systems.

\bibliographystyle{IEEEtran}
\bibliography{refs}

\begin{thebibliography}{10}
\providecommand{\url}[1]{#1}
\csname url@samestyle\endcsname
\providecommand{\newblock}{\relax}
\providecommand{\bibinfo}[2]{#2}
\providecommand{\BIBentrySTDinterwordspacing}{\spaceskip=0pt\relax}
\providecommand{\BIBentryALTinterwordstretchfactor}{4}
\providecommand{\BIBentryALTinterwordspacing}{\spaceskip=\fontdimen2\font plus
\BIBentryALTinterwordstretchfactor\fontdimen3\font minus \fontdimen4\font\relax}
\providecommand{\BIBforeignlanguage}[2]{{%
\expandafter\ifx\csname l@#1\endcsname\relax
\typeout{** WARNING: IEEEtran.bst: No hyphenation pattern has been}%
\typeout{** loaded for the language `#1'. Using the pattern for}%
\typeout{** the default language instead.}%
\else
\language=\csname l@#1\endcsname
\fi
#2}}
\providecommand{\BIBdecl}{\relax}
\BIBdecl

\bibitem{adavanne2018sound}
S.~Adavanne, A.~Politis, J.~Nikunen, and T.~Virtanen, ``Sound event localization and detection of overlapping sources using convolutional recurrent neural networks,'' \emph{IEEE Journal of Selected Topics in Signal Processing}, vol.~13, no.~1, pp. 34--48, 2019.

\bibitem{adavanne2019multi}
S.~Adavanne, A.~Politis, and T.~Virtanen, ``A multi-room reverberant dataset for sound event localization and detection,'' in \emph{Proc. of DCASE Workshop}, 2019.

\bibitem{politis2020dataset}
A.~Politis, S.~Adavanne, and T.~Virtanen, ``A dataset of reverberant spatial sound scenes with moving sources for sound event localization and detection,'' in \emph{Proc. of DCASE Workshop}, 2020.

\bibitem{politis2021dataset}
A.~Politis, S.~Adavanne, D.~Krause, A.~Deleforge, P.~Srivastava, and T.~Virtanen, ``A dataset of dynamic reverberant sound scenes with directional interferers for sound event localization and detection,'' in \emph{Proc. of DCASE Workshop}, 2021.

\bibitem{politis2020overview}
A.~Politis, A.~Mesaros, S.~Adavanne, T.~Heittola, and T.~Virtanen, ``Overview and evaluation of sound event localization and detection in dcase 2019,'' \emph{IEEE/ACM Transactions on Audio, Speech, and Language Processing}, vol.~29, pp. 684--698, 2021.

\bibitem{nguyen2022salsa}
T.~N.~T. Nguyen, D.~L. Jones, K.~N. Watcharasupat, H.~Phan, and W.-S. Gan, ``{SALSA-Lite}: A fast and effective feature for polyphonic sound event localization and detection with microphone arrays,'' in \emph{Proc. of IEEE ICASSP}, 2022, pp. 716--720.

\bibitem{mazzon2019first}
L.~Mazzon, Y.~Koizumi, M.~Yasuda, and N.~Harada, ``First order {Ambisonics} domain spatial augumentation for {DNN}-based direction of arrival estimation,'' in \emph{Proc. of DCASE Workshop}, 2019.

\bibitem{wang2023four}
Q.~Wang, J.~Du, H.-X. Wu, J.~Pan, F.~Ma, and C.-H. Lee, ``A four-stage data augmentation approach to resnet-conformer based acoustic modeling for sound event localization and detection,'' \emph{IEEE/ACM Transactions on Audio, Speech, and Language Processing}, vol.~31, pp. 1251--1264, 2023.

\bibitem{koyama2022spatial}
Y.~Koyama, K.~Shigemi, M.~Takahashi, K.~Shimada, N.~Takahashi, E.~Tsunoo, S.~Takahashi, and Y.~Mitsufuji, ``Spatial data augmentation with simulated room impulse responses for sound event localization and detection,'' in \emph{Proc. of IEEE ICASSP}, 2022, pp. 8872--8876.

\bibitem{kapka2019sound}
S.~Kapka and M.~Lewandowski, ``Sound source detection, localization and classification using consecutive ensemble of {CRNN} models,'' in \emph{Proc. of DCASE Workshop}, 2019.

\bibitem{cao2019polyphonic}
Y.~Cao, Q.~Kong, T.~Iqbal, F.~An, W.~Wang, and M.~D. Plumbley, ``Polyphonic sound event detection and localization using a two-stage strategy,'' in \emph{Proc. of DCASE Workshop}, 2019.

\bibitem{nguyen2020sequence}
T.~N. Tho~Nguyen, D.~L. Jones, and W.-S. Gan, ``A sequence matching network for polyphonic sound event localization and detection,'' in \emph{Proc. of IEEE ICASSP}, 2020, pp. 71--75.

\bibitem{cao2021improved}
Y.~Cao, T.~Iqbal, Q.~Kong, F.~An, W.~Wang, and M.~D. Plumbley, ``An improved event-independent network for polyphonic sound event localization and detection,'' in \emph{Proc. of IEEE ICASSP}, 2021.

\bibitem{shimada2021accdoa}
K.~Shimada, Y.~Koyama, N.~Takahashi, S.~Takahashi, and Y.~Mitsufuji, ``Accdoa: Activity-coupled cartesian direction of arrival representation for sound event localization and detection,'' in \emph{Proc. of IEEE ICASSP}, 2021, pp. 915--919.

\bibitem{politisstarss22}
A.~Politis, K.~Shimada, P.~Sudarsanam, S.~Adavanne, D.~Krause, Y.~Koyama, N.~Takahashi, S.~Takahashi, Y.~Mitsufuji, and T.~Virtanen, ``Starss22: A dataset of spatial recordings of real scenes with spatiotemporal annotations of sound events,'' in \emph{Proc. of DCASE Workshop}, 2022.

\bibitem{shimada2023starss23}
K.~Shimada, A.~Politis, P.~Sudarsanam, D.~A. Krause, K.~Uchida, S.~Adavanne, A.~Hakala, Y.~Koyama, N.~Takahashi, S.~Takahashi, T.~Virtanen, and Y.~Mitsufuji, ``Starss23: An audio-visual dataset of spatial recordings of real scenes with spatiotemporal annotations of sound events,'' \emph{Advances in neural information processing systems}, vol.~36, pp. 72\,931--72\,957, 2023.

\bibitem{aparicio2024baseline}
D.~D.-G. Aparicio, A.~Politis, P.~A. Sudarsanam, K.~Shimada, D.~Krause, K.~Uchida, Y.~Koyama, N.~Takahashi, S.~Takahashi, T.~Shibuya, Y.~Mitsufuji, and T.~Virtanen, ``Baseline models and evaluation of sound event localization and detection with distance estimation in dcase 2024 challenge,'' in \emph{Proc. of DCASE Workshop}, 2024, pp. 41--45.

\bibitem{Du_NERCSLIP_dcase23task3_report}
Q.~Wang, Y.~Jiang, S.~Cheng, M.~Hu, Z.~Nian, P.~Hu, Z.~Liu, Y.~Dong, M.~Cai, J.~Du, and C.-H. Lee, ``The nerc-slip system for sound event localization and detection of dcase2023 challenge,'' DCASE2023 Challenge, Tech. Rep., June 2023.

\bibitem{berghi2024fusion}
D.~Berghi, P.~Wu, J.~Zhao, W.~Wang, and P.~J.~B. Jackson, ``Fusion of audio and visual embeddings for sound event localization and detection,'' in \emph{Proc. of IEEE ICASSP}, 2024, pp. 8816--8820.

\bibitem{Du_NERCSLIP_dcase22task3_report}
Q.~Wang, L.~Chai, H.~Wu, Z.~Nian, S.~Niu, S.~Zheng, Y.~Wang, L.~Sun, Y.~Fang, J.~Pan, J.~Du, and C.-H. Lee, ``The nerc-slip system for sound event localization and detection of dcase2022 challenge,'' DCASE2022 Challenge, Tech. Rep., June 2022.

\bibitem{Hu_IACAS_dcase22task3_report}
J.~Hu, Y.~Cao, M.~Wu, Q.~Kong, F.~Yang, M.~D. Plumbley, and J.~Yang, ``Sound event localization and detection for real spatial sound scenes: Event-independent network and data augmentation chains,'' DCASE2022 Challenge, Tech. Rep., June 2022.

\bibitem{partha2021dcase}
P.~Sudarsanam, A.~Politis, and K.~Drossos, ``Assessment of self-attention on learned features for sound event localization and detection,'' in \emph{Proc. of DCASE Workshop}, 2021.

\bibitem{Du_NERCSLIP_dcase24task3_report}
Q.~Wang, Y.~Dong, H.~Hong, R.~Wei, M.~Hu, S.~Cheng, Y.~Jiang, M.~Cai, X.~Fang, and J.~Du, ``The nerc-slip system for sound event localization and detection with source distance estimation of dcase 2024 challenge,'' DCASE2024 Challenge, Tech. Rep., June 2024.

\bibitem{shimada2024savgbench}
K.~Shimada, C.~Simon, T.~Shibuya, S.~Takahashi, and Y.~Mitsufuji, ``Savgbench: Benchmarking spatially aligned audio-video generation,'' \emph{arXiv preprint arXiv:2412.13462}, 2024.

\bibitem{krause2024sound}
D.~A. Krause, A.~Politis, and A.~Mesaros, ``Sound event detection and localization with distance estimation,'' in \emph{Proc. of EUSIPCO}, 2024, pp. 286--290.

\bibitem{krause2023binaural}
D.~A. Krause, G.~Garc{\'\i}a-Barrios, A.~Politis, and A.~Mesaros, ``Binaural sound source distance estimation and localization for a moving listener,'' \emph{IEEE/ACM Transactions on Audio, Speech, and Language Processing}, vol.~32, pp. 996--1011, 2023.

\bibitem{kushwaha2023sound}
S.~S. Kushwaha, I.~R. Roman, M.~Fuentes, and J.~P. Bello, ``Sound source distance estimation in diverse and dynamic acoustic conditions,'' in \emph{Proc. of IEEE WASPAA}, 2023, pp. 1--5.

\bibitem{liang2023reconstructing}
B.~S. Liang, A.~S. Liang, I.~Roman, T.~Weiss, B.~Duinkharjav, J.~P. Bello, and Q.~Sun, ``Reconstructing room scales with a single sound for augmented reality displays,'' \emph{Journal of Information Display}, vol.~24, no.~1, pp. 1--12, 2023.

\bibitem{wilkins2023two}
J.~Wilkins, M.~Fuentes, L.~Bondi, S.~Ghaffarzadegan, A.~Abavisani, and J.~P. Bello, ``Two vs. four-channel sound event localization and detection,'' in \emph{Proc. of DCASE Workshop}, 2023.

\bibitem{he2016deep}
K.~He, X.~Zhang, S.~Ren, and J.~Sun, ``Deep residual learning for image recognition,'' in \emph{Proc. of IEEE CVPR}, 2016, pp. 770--778.

\bibitem{vaswani2017attention}
A.~Vaswani, N.~Shazeer, N.~Parmar, J.~Uszkoreit, L.~Jones, A.~N. Gomez, {\L}.~Kaiser, and I.~Polosukhin, ``Attention is all you need,'' \emph{Advances in neural information processing systems}, vol.~30, 2017.

\bibitem{shimada2022multi}
K.~Shimada, Y.~Koyama, S.~Takahashi, N.~Takahashi, E.~Tsunoo, and Y.~Mitsufuji, ``Multi-accdoa: Localizing and detecting overlapping sounds from the same class with auxiliary duplicating permutation invariant training,'' in \emph{Proc. of IEEE ICASSP}, 2022, pp. 316--320.

\bibitem{roman2024spatial}
I.~R. Roman, C.~Ick, S.~Ding, A.~S. Roman, B.~McFee, and J.~P. Bello, ``Spatial scaper: a library to simulate and augment soundscapes for sound event localization and detection in realistic rooms,'' in \emph{Proc. of IEEE ICASSP}, 2024, pp. 1221--1225.

\bibitem{fonseca2021fsd50k}
E.~Fonseca, X.~Favory, J.~Pons, F.~Font, and X.~Serra, ``Fsd50k: an open dataset of human-labeled sound events,'' \emph{IEEE/ACM Transactions on Audio, Speech, and Language Processing}, vol.~30, pp. 829--852, 2021.

\bibitem{defferrard2017fma}
M.~Defferrard, K.~Benzi, P.~Vandergheynst, and X.~Bresson, ``Fma: A dataset for music analysis,'' in \emph{Proc. of ISMIR}, 2017.

\bibitem{roman2025generating}
A.~S. Roman, A.~Chang, G.~Meza, and I.~R. Roman, ``Generating diverse audio-visual 360 soundscapes for sound event localization and detection,'' \emph{arXiv preprint arXiv:2504.02988}, 2025.

\end{thebibliography}

%
%
%
%
%
%
%
%
%

\end{document}